\documentclass{Interspeech}
\usepackage{graphicx}
\usepackage{subcaption}
\usepackage{stfloats}
\usepackage{float}
\usepackage{multirow}
\usepackage{algorithm}
\usepackage{algorithmicx}
\usepackage{longtable}
\usepackage{tabularray}
\usepackage{colortbl}
\usepackage{xcolor}
\usepackage{listings}
\usepackage{graphicx}
\usepackage{subcaption}
 
\lstdefinestyle{lfonts}{
  basicstyle   = \footnotesize\ttfamily,
  stringstyle  = \color{purple},
  keywordstyle = \color{blue!60!black}\bfseries,
  commentstyle = \color{olive}\scshape,
}
\lstdefinestyle{lnumbers}{
  numbers     = left,
  numberstyle = \tiny,
  numbersep   = 1em,
  firstnumber = 1,
  stepnumber  = 1,
}
\lstdefinestyle{llayout}{
  breaklines       = true,
  tabsize          = 2,
  columns          = flexible,
}
\lstdefinestyle{lgeometry}{
  xleftmargin      = 20pt,
  xrightmargin     = 0pt,
  frame            = tb,
  framesep         = \fboxsep,
  framexleftmargin = 20pt,
}
\lstdefinestyle{lgeneral}{
  style = lfonts,
  style = lnumbers,
  style = llayout,
  style = lgeometry,
}
\lstdefinestyle{python}{
    language = {Python},
    style    = lgeneral,
}
\lstdefinestyle{lnoNumbers}{
  style = lfonts,
  style = llayout,
  style = lgeometry,
}

\makeatletter
\let\IS@orig@affiliation\affiliation         % save original

\makeatletter
\renewcommand\anonaddress{%
  \ifinterspeechfinal
    {\centering
      \vskip-3pt
      \shortstack[c]{%
        \textsuperscript{1}Zhejiang University, China\quad
        \textsuperscript{2}UC Berkeley, United States\\
        \textsuperscript{3}UCSF, United States
      }\par
      \vskip-3pt
    }%
  \fi
}
\makeatother



\interspeechcameraready

\title{Dysfluent WFST: A Framework for Zero-Shot Speech Dysfluency Transcription and Detection}

\author[affiliation={ 1 }]{Chenxu}{Guo}
\author[affiliation={ 2 }]{Jiachen}{Lian}
\author[affiliation={ 1 }]{Xuanru}{Zhou}
\author[affiliation={ 1 }]{Jinming}{Zhang}
\author[affiliation={ 1 }]{Shuhe}{Li}
\author[affiliation={ 1 }]{Zongli}{Ye}
\author[affiliation={ 2 }]{Hwi Joo}{Park}
\author[affiliation={ 2 }]{Anaisha}{Das}
\author[affiliation={ 3 }]{Zoe}{Ezzes}
\author[affiliation={ 3 }]{Jet}{Vonk}
\author[affiliation={ 3 }]{Brittany}{Morin}
\author[affiliation={ 3 }]{Rian}{Bogley}
\author[affiliation={ 3 }]{Lisa}{Wauters}
\author[affiliation={ 3 }]{Zachary}{Miller}
\author[affiliation={ 3 }]{Maria}{Gorno-Tempini}
\author[affiliation={ 2 }]{Gopala}{Anumanchipalli}

\affiliation{}{Zhejiang University}{China}
\affiliation{}{University of California, Berkeley}{United States}
\affiliation{}{University of California, San Francisco}{United States}
\email{louis.kwok.work@gmail.com, jiachenlian@berkeley.edu, gopala@berkeley.edu}

\keywords{speech recognition, dysfluency detection}

\usepackage{comment}

\begin{document}

\maketitle

\begin{abstract}
% \textcolor{red}{[Please rewrite abstract based on introduction]}
Automatic detection of speech dysfluency aids speech-language pathologists in efficient transcription of disordered speech, enhancing diagnostics and treatment planning. Traditional methods, often limited to classification, provide insufficient clinical insight, and text-independent models misclassify dysfluency, especially in context-dependent cases. This work introduces Dysfluent-WFST, a zero-shot decoder that simultaneously transcribes phonemes and detects dysfluency. Unlike previous models, Dysfluent-WFST operates with upstream encoders like WavLM and requires no additional training. It achieves state-of-the-art performance in both phonetic error rate and dysfluency detection on simulated and real speech data. Our approach is lightweight, interpretable, and effective, demonstrating that explicit modeling of pronunciation behavior in decoding, rather than complex architectures, is key to improving dysfluency processing systems. Our implementation is open-source at \href{https://github.com/Berkeley-Speech-Group/DysfluentWFST}{https://github.com/Berkeley-Speech-Group/DysfluentWFST}.
\end{abstract}

\vspace{-2.5mm}
\section{Introduction}

Automatic detection of speech dysfluency is a key module that helps speech-language pathologists efficiently transcribe disordered speech, which then serves as strong evidence for diagnostic purposes or even for treatment feedback planning. For a long time, research in this area has been constrained to a sentence classification task: determining whether a given speech sample contains stuttering or not~\cite{HMM,barrett2022systematic-stutter1.0,jouaiti2022dysfluency-stutter1.1,bayerl2022detecting-stutter1.2,zayats2016disfluency-stutter1.3,montacie2022audio-stutter1.4,kourkounakis2021fluentnet, alharbi2017segment-detection2, alharbi2020segment-detection3, segment-detection4, shonibare2022frame-detection2, wagner2024largelanguagemodelsdysfluency, shih2024self-ssl-stutter}. However, binary or categorical classification results provide limited information for real clinical applications. Additionally, text-independent modeling of stuttered speech can be problematic, as certain cases, such as "I saw a dodo," may be misclassified as stuttering of "do". Without textual context, the aforementioned methods are likely to misdetect such instances as dysfluent speech. Therefore, the detection of dysfluent speech is text-dependent, which was ignored in previous work. 

UDM~\cite{UDM} shifted this paradigm by focusing on a more challenging and general task: transcribing the specific type of dysfluency (e.g., repetition, insertion, replacement, deletion, prolongation) along with its corresponding timestamps given a dysfluent speech sample and its text. Specifically, UDM~\cite{UDM} and its hierarchical extension~\cite{lian-anumanchipalli-2024-towards} stated that two modules are essential: a \textit{transcription module} that accurately transcribes the verbatim speech and a \textit{detection module} that identifies dysfluencies based on the transcription. The limitations of these works are evident. The entire system is rule-driven, making it non-scalable with respect to data. Some subsequent research efforts have attempted to enhance this pipeline. For example, YOLO-Stutter~\cite{Zhou2024YOLOStutterER} integrated two modules into an end-to-end pipeline, which was further extended to a multilingual setting~\cite{zhou2024stutter}. However, the primary limitation of these end-to-end models is that their performance remains only marginally better than rule-based approaches, and scaling up remains challenging due to a lack of interpretability. To address these issues, SSDM~\cite{ssdm} introduced an LCS alignment module that ensures interpretability while maintaining a fully differentiable system with LLM integration, establishing itself as the current state-of-the-art pipeline. However, SSDM~\cite{ssdm} and its extension~\cite{lian2024ssdm2.0} suffer from a highly complex architectural design and do not effectively address phonetic dysfluency transcription, which we identify as the critical bottleneck for this problem.

In fact, \textit{detecting dysfluency would not be particularly challenging if an accurate verbatim phonetic transcription could be obtained}. For example, if a model transcribes "She's n-not (N AA N AA T) here." while the ground truth is "She's not here," then detection becomes straightforward—one could develop alignment rules~\cite{ssdm} or simply call an LLM API~\cite{OpenAI_chatgpt}. Notably, LLM-driven speech dysfluency detection methods~\cite{wagner2024largelanguagemodelsdysfluency, ssdm, lian2024ssdm2.0} leverage the power of LLMs to enhance performance. However, due to the poor quality of phonetic transcription, their performance remains upper-bounded. Therefore, \textit{our goal is to provide more accurate phonetic transcription for dysfluent speech.} 

In state-of-the-art phoneme recognition systems~\cite{li2020universal,xu2021simple-w2v2-phoneme,wavlm-ctc}, phoneme transcription is typically performed using greedy search or beam search on the emission matrix. While constrained decoding~\cite{graves2006connectionist-ctc} assumes strict monotonic alignment between speech and text, dysfluent speech often follows a non-monotonic alignment. In unconstrained settings, where the ground truth is unknown, applying greedy or unconstrained Viterbi search~\cite{UDM} can introduce random phoneme insertions or jumps. Dysfluent speech, however, is shaped by human pronunciation behavior—specific phonemes or syllables are affected, and insertions or repetitions occur systematically. Thus, by incorporating pronunciation priors into decoding, we can achieve more accurate phoneme transcription.

In this work, we introduce \textit{Dysfluent-WFST}, a zero-shot decoder that operates seamlessly with upstream encoders such as wav2vec2~\cite{Baevski2020wav2vec2A} or WavLM~\cite{chen2022wavlm}. \textit{Dysfluent-WFST} accept emissions from encoder and simultaneously generates both phoneme transcription and dysfluency detection results. Notably, \textit{Dysfluent-WFST} is lightweight and requires no additional training, being parameter-free. Despite its simplicity, it achieves \textit{state-of-the-art phonetic error rate (PER)} in our dysfluent phoneme recognition benchmark.
As dysfluency detection occurs simultaneously, \textit{Dysfluent-WFST} also achieves state-of-the-art performance in dysfluency detection on both simulated and real speech benchmarks. While one related work~\cite{Kouzelis2023WeaklysupervisedFA} also utilized WFST to achieve better forced alignment in dysfluent speech, their approach focuses on aligning dysfluent speech without manual transcription and has not explored core dysfluency detection problems: transcription and detection. To our knowledge, \textit{Dysfluent-WFST} is the most lightweight, interpretable, and effective speech dysfluency transcriber.

\begin{figure}[t]
    \centering
    \includegraphics[width=1.0\linewidth]{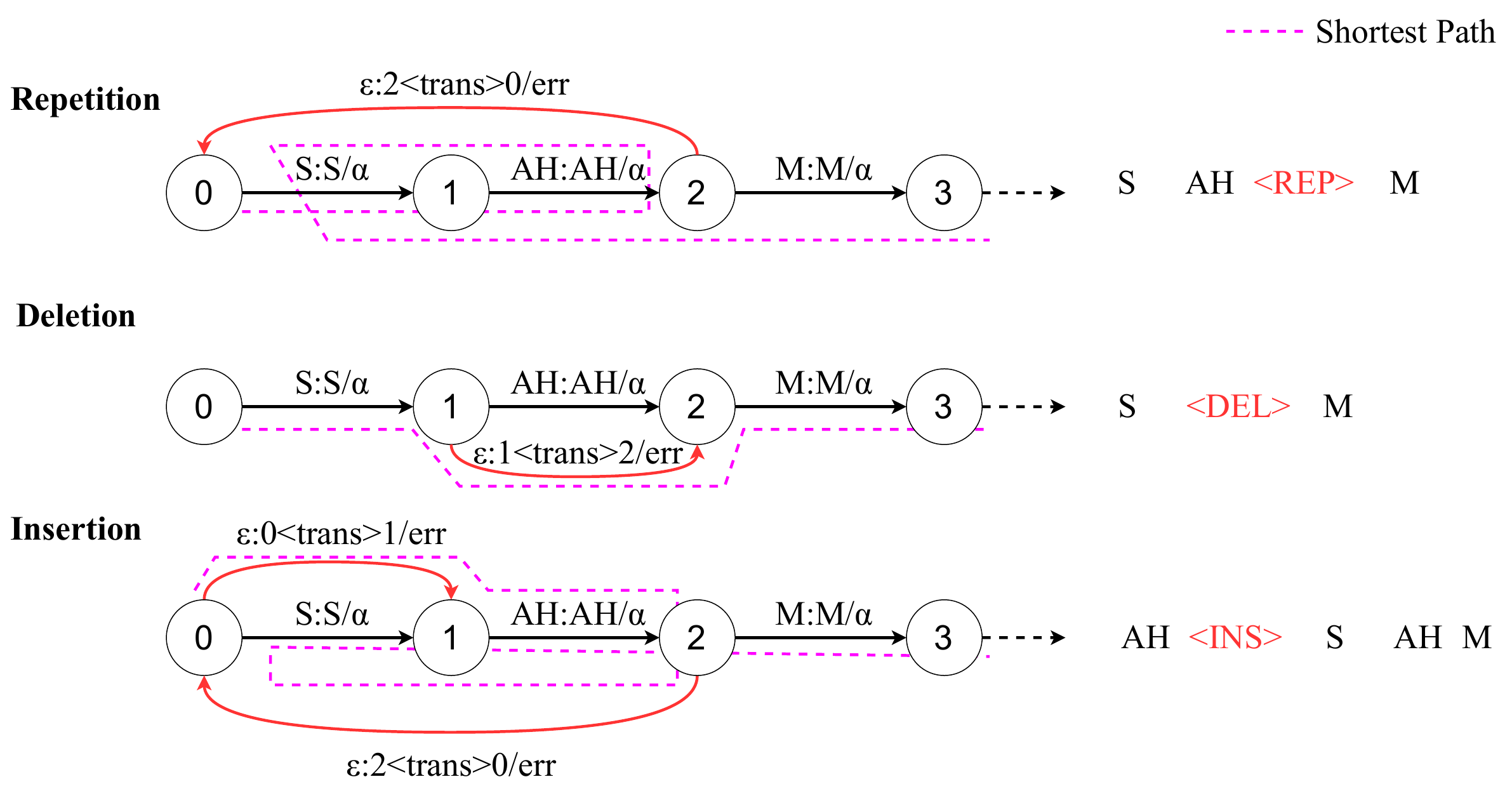}
    \caption{Illustrating Dysfluency Transitions: Repetition, Deletion, and Insertion in WFST}
    \label{fig:exp}
\end{figure}

\begin{figure*}[htbp]
    \centering
    \includegraphics[width=1.0\linewidth]{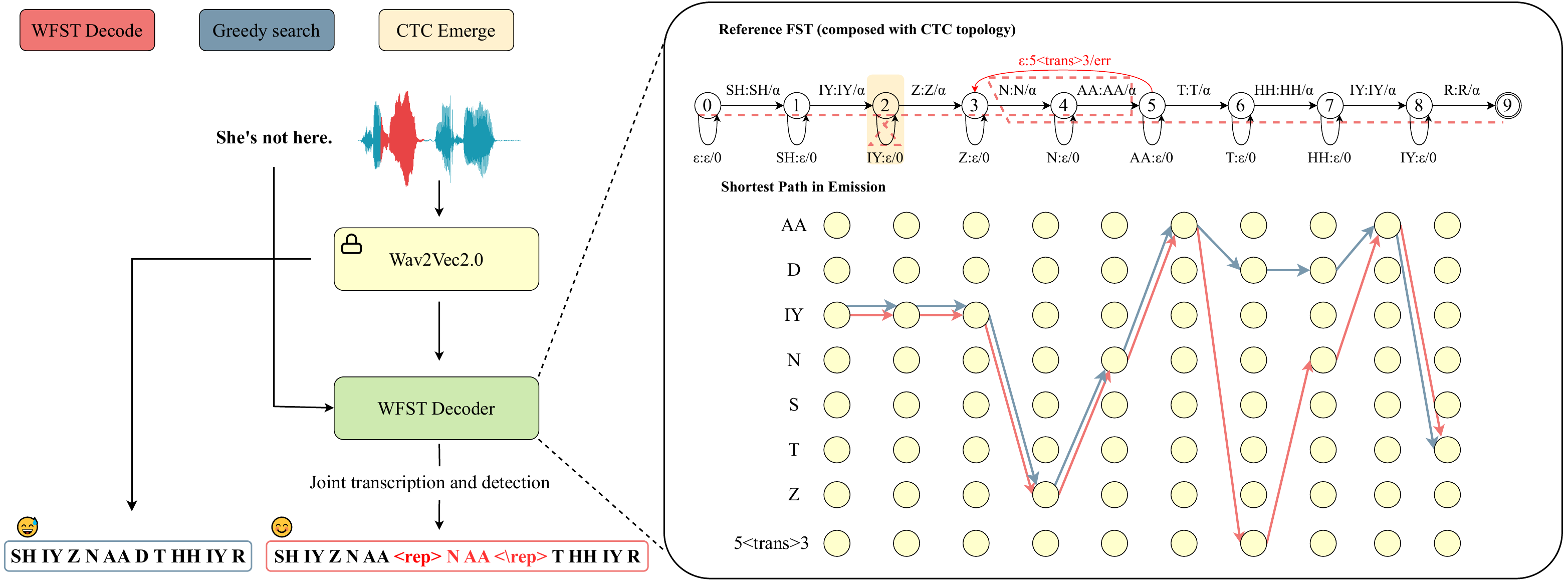}
    \caption{Decoder Workflow Based on WFST: The framework takes as input the speech signal and the corresponding reference text, and outputs the phoneme transcription sequence along with dysfluency detection (e.g., repetition). In this example, the reference text is "She's not here," while the spoken audio is "She's n-not (N AA N AA T) here." When using greedy search, the model may incorrectly force-align the repetition part to non-monotonic phonemes, such as "D." In contrast, our WFST-based method incorporates a return arc, enabling the repetition of phonemes present in the reference text. As the shortest path traverses this return arc, the output is labeled as "5<trans>3," indicating a repetition in speech. Consequently, our WFST-based decoder successfully outputs the correct phoneme transcription along with accurate dysfluency detection.
    }
    \label{fig:workflow}
\end{figure*}

\vspace{-5mm}
\section{Method}

A \textit{weighted finite-state acceptor} (WFSA) is a 6-tuple \( \mathcal{A} = (\Sigma, \mathbb{Q}, \mathbb{Q}_s, \mathbb{Q}_a, \pi, \omega) \), where \( \Sigma \) is the input alphabet, \( \mathbb{Q} \) is the set of states, \( \mathbb{Q}_s \subseteq \mathbb{Q} \) and \( \mathbb{Q}_a \subseteq \mathbb{Q} \) are the sets of start and accepting states, respectively. The transition function \( \pi: \mathbb{Q} \times \Sigma \to \mathbb{Q} \) defines state transitions, while the weight function \( \omega: \mathbb{Q} \times \Sigma \to \mathbb{R} \) assigns a weight to each transition. 

A \textit{weighted finite-state transducer} (WFST) extends WFSA by incorporating an output alphabet \( \Delta \), forming a 7-tuple \( \mathcal{T} = (\Sigma, \Delta, \mathbb{Q}, \mathbb{Q}_s, \mathbb{Q}_a, \pi, \omega) \), where \( \pi: \mathbb{Q} \times \Sigma \times \Delta \to \mathbb{Q} \) maps transitions, and \( \omega: \mathbb{Q} \times \Sigma \times \Delta \to \mathbb{R} \) assigns weights. Each edge in the WFST connects two states and is labeled with an input \( p \in \Sigma \), an output \( r \in \Delta \), and a weight \( w \in \mathbb{R} \).

For the methods described in this Section, we assume: \textbf{(a)} The input sequence \( X = \left[x_0, x_1, \dots, x_{T-1} \right] \) represents a speech signal, \( T \) is the number of time steps. \textbf{(b)} The reference phoneme sequence is denoted as \( R = \left[r_0, r_1, \dots, r_{S-1} \right] \), where \( S \) is the number of phonemes in the sequence. \textbf{(c)} The emission tensor \( \Xi_x \in \mathbb{R}^{T \times C} \) where \( C \) is the number of phoneme classes. \textbf{(d)} The phonetic transcription is represented as \( Y = \left[y_0, y_1, \dots, y_{N-1} \right] \), where \( N \) is the number of phonetic labels.
\vspace{-5pt}
\subsection{Proposed workflow}

We can observe the detailed transition process during decoding, as shown in Figure~\ref{fig:exp}, where we apply following rules to capture dysfluencies: (a) \textbf{Regular}: Speech without any dysfluency follows a horizontal linear path, with continuously increasing indices; (b) \textbf{Repetition}: When the path revisits a state that has occurred previously, it indicates a repetition; (c) \textbf{Deletion}: When the path skips a state and proceeds to a state with a larger index, it signifies a deletion; (d) \textbf{Insertion}: When the shortest path passes through a state with a smaller index, it is classified as an insertion.

 Our framework takes the reference text and corresponding speech as input.
 Using WFST-based alignment, we match the phonemes spoken by the speaker with the expected phonemes in the reference text.
Figure~\ref{fig:workflow} illustrates the entire pipeline, consider the example where the speaker utters "She’s n-not (N AA N AA T) here," with the ground truth being "She's not here." Traditional methods, such as beam search, may produce non-monotonic alignments (e.g., outputting "D" as shown). In contrast, our WFST-based approach maintains meaningful state transitions. When the decoding path traverses a return arc, it corresponds to the detection of a "repetition," as illustrated in Figure~\ref{fig:exp}.
\vspace{-5pt}
\begin{lstlisting}[style = python, caption={Phoneme Recognition and Dysfluency Detection}, label={algo:1}]
# input: reference text and audio file
ref_text = "She's not here"
audio = "demo.wav"

# encode, provide encoder lexcion
model = speech2emision()
emission = logits = infer(model, audio, lexicon)['logits']

# decode
# first, build ref_fst, ctc_topo, and emission graph
ref_fst = ref_fst(ref_text.to_phoneme())
ctc_fsa = ctc_topo(len(lexicon))
emission_fsa = emission_fsa(emission)

# composed and intersect
composed = compose(ctc_fsa, ref_fst, treat_epsilons_specially=True)
lattices = intersect(composed, emission_fsa)
shortest = shortest_path(lattices)
phoneme_seq = [self.lexcion[i] for i in shortest]

# clean_states: [(1, 2, 'SH'), ..., (start_state, end_state, phoneme)]
clean_states = extract_phnlst(phn_seq)

# detect dysfluency
def detect_dysfluency(phn_seq):
    # Detect dysfluency types
    for start, end, phoneme in clean_states:
        min_time = min(state_history)
        if start in state_history:
            dysfluency_type = "repetition"
        elif start < min_time:
            dysfluency_type = "insertion"
        elif start > prev_end + 1:
            dysfluency_type = "deletion"
        else:
            dysfluency_type = "normal"
        dysfluency_results.append({
        "phoneme": phoneme, 
        "dysfluency_type": dysfluency_type
        })
        state_history.add(start)
        prev_end = end
    return dysfluency_results
    
output = detect_dysfluency(clean_states)
\end{lstlisting}

As shown in Algorithm~\ref{algo:1}, we first use a speech encoder~\cite{Baevski2020wav2vec2A, chen2022wavlm} to generate an emission matrix \( \Xi_x \in \mathbb{R}^{T \times C} \), representing the log posterior probabilities of phoneme classes over \( T \) frames of speech. This matrix is passed to the decoder, which combines a Weighted Finite-State Automaton (WFSA) and Weighted Finite-State Transducer (WFST).

The reference text is used to construct a reference FST, which, unlike traditional ASR relying solely on FSAs, accounts for dysfluencies like insertions, deletions, and repetitions by extending the FSA into a WFST. This WFST allows the recognition of dysfluency types through return and skip arcs.

Next, we construct the emission graph \( \mathcal{\xi}_x \) using phone posterior probabilities from \( \Xi_x \), with edge weights corresponding to \( \log P(p_t | X) \). Using the phoneme class size \( C \), we build the CTC topology graph \( \mathcal{T} \) and a novel FST topology \( \mathcal{S} \). The decoding graph is then formed as:

\[
    \mathcal{D} = \left( \mathcal{T} \circ \mathcal{S} \right) \cap \mathcal{\xi}_x
\]

Finally, the optimal decoded phoneme sequence and dysfluency detection results are obtained by computing the shortest path in \( \mathcal{D} \). All WFST operations are performed using the k2.\footnote{https://github.com/k2-fsa/k2}.

\vspace{-5pt}
\subsection{Weight selection}

We introduce a parameter \( \beta \) to quantify speech dysfluency severity. For each state, we assign a weight \( \alpha = 1 - 10^{-\beta} \) to transitions along the horizontal arc following the expected transcription. This gives each phoneme a base error rate \( \text{err}_0 = 10^{-\beta} \) for transitions along return and skip arcs.

Building upon \cite{Kouzelis2023WeaklysupervisedFA}, we introduce a dynamic weighting mechanism. Instead of using a fixed error rate, we define the error \( \text{err} \) for non-forward paths. Intuitively, transitions to "distant" phonemes (in the reference sequence) are assigned lower weights or probabilities. The error is dynamically computed as:

\[
    \text{err} = \text{err}_0 \frac{1}{\sqrt{2\pi}} e^{-\frac{x^2}{2}},
\]

where \( x \) is the \( L_1 \) distance between phonemes \( r_i \) and \( r_j \) in the reference sequence, \( |i - j| \). This dynamic weighting adjusts transition probabilities based on phoneme distance, leading to a more context-aware and smoother decoding process. The parameter \( \beta \) controls the base error rate \( \text{err}_0 \), with larger values indicating more severe dysfluencies.

\begin{table*}[htbp]
\centering
\caption{Evaluation of Phoneme Recognition Across Different Frameworks. We applied W2V2 and WavLM-CTC as encoders, and compared the performance of greedy search and WFST-based decoders on various types of simulated data and nfvPPA. 
}
\label{tab:1}
\begin{tabular}{l|llll|llll} 
\hline
\multirow{2}{*}{Method} & \multicolumn{4}{c|}{PER}                        & \multicolumn{4}{c}{WPER}                         \\

                        & simu-rep        & simu-del & simu-ins & nfvPPA  & simu-rep        & simu-del & simu-ins & nfvPPA   \\ 
\hline\hline
W2V2-Greedy Search     & 34.13\%                  & 37.34\%                  & 32.24\%                  & 68.70\% & 22.83\%                  & 24.64\%                  & 20.98\%                  & 46.25\% \\
WavLMCTC-Greedy Search & 19.37\%                  & 21.33\%                  & 19.80\%                  & 61.49\% & 14.23\%                  & 15.21\%                  & 14.30\%                  & 49.95\% \\
W2V2-WFST (Ours)              & \textbf{\textbf{9.95\%}} & 13.49\%                  & 7.28\%                   & 55.37\% & \textbf{\textbf{9.34\%}} & 12.52\%                  & 6.21\%                   & 39.94\% \\
WavLMCTC-WFST (Ours)         & 10.71\%                  & \textbf{\textbf{4.46\%}} & \textbf{\textbf{2.80\%}} & \textbf{44.75\%} & 10.19\%                  & \textbf{\textbf{4.03\%}} & \textbf{\textbf{2.76\%}} & \textbf{32.20\%} \\
\hline
\end{tabular}
\end{table*}

\vspace{-2.5mm}
\section{Experiments}

\subsection{Dataset} \label{sec: datasets}

\textbf{(a) PPA Dataset}~\cite{gorno2011classification-ppa}: The nfvPPA dataset consists of speech samples from 35 individuals diagnosed with Nonfluent/Agrammatic Primary Progressive Aphasia. For evaluation purposes, we manually annotated the phonemes using Praat and Adobe Audition, totally 1h1min12s. \textbf{(b) Simulated Dataset}: To evaluate model performance, we created a synthetic speech dataset by leveraging Large Language Models. We provided \texttt{claude-3-5-sonnet-20241022}~\cite{Enis2024FromLT} with clean text and its corresponding CMU/IPA sequences, instructing it to insert dysfluencies at positions it deemed naturalistic. This process yielded both dysfluent IPA sequences (used for subsequent VITS~\cite{Kim2021ConditionalVA} speech synthesis) and CMU sequences with dysfluency labels (matching our model's output format for evaluation purposes). The resulting synthesized speech demonstrates high naturalness and authenticity, showing significant advantages over traditional rule-based methods for generating dysfluent IPA sequences~\cite{zhou2024timetokensbenchmarkingendtoend}. For different dysfluency types, we have generated three one-hour datasets: simulated repetition (simu-rep), simulated deletion (simu-del), and simulated insertion (simu-ins).

\vspace{-5pt}
\subsection{Evaluation metrics}

In phoneme recognition tasks, we replace the traditional Phonetic Error Rate (PER) with the \textit{Weighted Phonetic Error Rate} (WPER), which offers a more context-sensitive evaluation. Unlike PER, which treats all phonemes as equidistant, WPER accounts for phonetic variations by incorporating phoneme similarities, addressing the limitations of standard error metrics and providing a more accurate assessment of recognition performance.
To define WPER, we first construct a similarity matrix\footnote{https://github.com/Berkeley-Speech-Group/DysfluentWFST} \(S\) of size \(39 \times 39\), representing the similarity between each pair of CMU phonemes (excluding lexical stress). 
The WPER is then defined as:

\begin{equation}
    \text{WPER} = \frac{\displaystyle \sum^{(i, j)}( 1-S(i, j))\ + D + I}{N}
\end{equation}

where $i$, $j$ denotes target phoneme and substitute phoneme, \(S(i, j)\) represents the similarity score between phoneme \(i\) and \(j\), $N$ is the length of reference phoneme sequence, and \(D\), \(I\) account for deletion and insertion errors, respectively. In the case of a substitution, the phoneme similarity is considered first before applying the error penalties.

\vspace{-5pt}
\subsection{Results and discussions}

To evaluate the dysfluency transcription performance of the WFST decoder, we conduct experiments on both simulated data, which includes various dysfluency types (repetition, deletion, and insertion), and the nfvPPA dataset, which is more representative of real-world scenarios. The evaluation metrics used are the Phonetic Error Rate (PER) and the Weighted Phonetic Error Rate (WPER). Furthermore, we compare the performance of several encoders, including state-of-the-art models such as Wav2Vec 2.0 and WavLM. For each encoder, we compare the results obtained using the default decoding methods with those produced by our proposed WFST-based approach. The evaluation results are presented in Table~\ref{tab:1}.

For dysfluency detection, we evaluated YOLO-Stutter on simulated dataset. The accuracy is approximately 50\% for repetition detection and only 5\% for deleted phonemes. In contrast, our WFST-based approach demonstrated strong performance without any prior training. Specifically, for repetition detection, WFST achieved \textbf{100\%} accuracy at the count level. This suggests that the structured nature of WFST effectively captures repetition patterns in dysfluent speech, making it a robust candidate for dysfluency detection tasks. Our findings highlight the effectiveness of WFST in zero-shot dysfluency detection, particularly for repetition patterns, while also emphasizing the need for fair benchmarking across different models.
\vspace{-7pt}
\begin{table}[htbp]
\centering
\caption{Dysfluency Detection Accuracy Evaluation (\%)}
\label{tab:3}
\vspace{-5pt}
\setlength{\tabcolsep}{6pt} 
\renewcommand{\arraystretch}{0.8} 
\resizebox{6cm}{!}{
\begin{tblr}{
  row{1} = {c},
  cell{1}{1} = {r=2}{},
  cell{1}{2} = {c=3}{},
  vline{2} = {1-5}{},
  hline{1,6} = {-}{},
  hline{3} = {1-4}{},
}
Method       & Accuracy   &           &           \\
             & simu-rep & simu-del & simu-ins \\
W2V2-WFST    & 100       & 50  & 12   \\
WavLM-WFST   & 100       & 8   & 34  \\
Yolo-stutter & 51        & 5        & -         
\end{tblr}}
\end{table}

We observe that WFST decoding performance is highly sensitive to the emission. To investigate this, we add Gaussian noise \( n_i \sim \mathcal{N}(0, \sigma^2) \) to the emission and denote \( \sigma \) as the noise level. The results, shown in Table~\ref{tab:4}, indicate that WFST decoding lacks robustness: even small amounts of noise in the emission significantly alter the transcription.

\begin{table}
\centering
\caption{WPER(\%) when emission add different levels of noise (test data: simu-rep)}
\label{tab:4}
\vspace{-5pt}
\resizebox{8cm}{!}{
\begin{tblr}{
  vline{2} = {-}{},
  hline{1-2,6} = {-}{},
}
Noise level~\(\sigma\)            & 0 & 0.1 & 1 & 10 & 100 \\
W2V2-Greedy Search     & 22.83 & 22.68   & 24.40 & 55.23  & 484.9   \\
WavLMCTC-Greedy Search & 14.23 & 15.37   & 18.96 & 13.18  & 13.36   \\
W2V2-WFST              & 9.43 & 74.06   & 76.09 & 142.27  & 130.67   \\
WavLMCTC-WFST               & 10.19 & 29.77   & 31.42 & 33.27  & 102.33   
\end{tblr}}
\end{table}

\subsection{Ablation experiments}
\begin{figure}[h]
    \centering
    \includegraphics[width=0.88\linewidth]{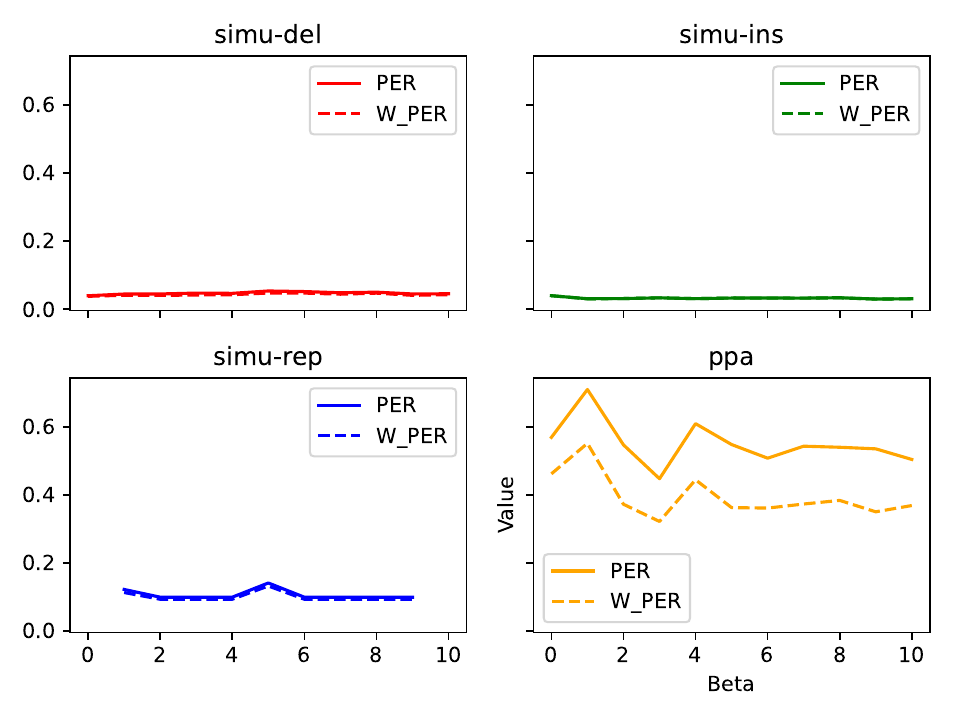}
    \caption{Impact of varying \( \beta \) values on transcription performance on different dataset}
    \label{fig:3}
\end{figure}

We observe that within a certain threshold, increases in \( \beta \) yield negligible improvements in the results. However, when \( \beta \) becomes excessively large, the modified FST may degrade into a linear structure due to limitations in floating-point precision within the Python environment (Figure~\ref{fig:3}).

\vspace{-5pt}

\vspace{-5mm}
\section{Conclusion and future work}

Our decoder achieves strong performance in dysfluency transcription, with a low phonetic error rate (PER) despite not requiring dysfluent speech in training. The architecture is interpretable and efficient. While it excels at detecting repetitions, its accuracy for insertions and deletions is lower.
The WFST decoding approach finds the shortest path but struggles with insertions and deletions, which require skipping edges. These edges, despite having higher costs, may be selected for their ability to quickly reach the endpoint, making their overall weight comparable to that of the expected path.
Additionally, WFST’s lack of a differentiable mathematical representation complicates training. Future work will focus on improving the handling of skipping edges and exploring joint training of the WFST and encoder. It also is worth exploring articulatory pronunciation feedback in kinematics space~\cite{cho2024jstsp, wu23k_interspeech} or gestural space~\cite{ssdm, lian22bcsnmf, lian2023factor} with closed-loop dysfluency correction.

\section{Acknowledgements}
Thanks for support from UC Noyce Initiative, Society of Hellman Fellows, NIH/NIDCD, and the Schwab Innovation fund.

% \ifinterspeechfinal
%      The Interspeech 2025 organisers
% \else
%      The authors
% \fi
% would like to thank ISCA and the organising committees of past Interspeech conferences for their help and for kindly providing the previous version of this template.

\bibliographystyle{IEEEtran}
\bibliography{mybib}

\begin{thebibliography}{10}
\providecommand{\url}[1]{#1}
\csname url@samestyle\endcsname
\providecommand{\newblock}{\relax}
\providecommand{\bibinfo}[2]{#2}
\providecommand{\BIBentrySTDinterwordspacing}{\spaceskip=0pt\relax}
\providecommand{\BIBentryALTinterwordstretchfactor}{4}
\providecommand{\BIBentryALTinterwordspacing}{\spaceskip=\fontdimen2\font plus
\BIBentryALTinterwordstretchfactor\fontdimen3\font minus \fontdimen4\font\relax}
\providecommand{\BIBforeignlanguage}[2]{{%
\expandafter\ifx\csname l@#1\endcsname\relax
\typeout{** WARNING: IEEEtran.bst: No hyphenation pattern has been}%
\typeout{** loaded for the language `#1'. Using the pattern for}%
\typeout{** the default language instead.}%
\else
\language=\csname l@#1\endcsname
\fi
#2}}
\providecommand{\BIBdecl}{\relax}
\BIBdecl

\bibitem{HMM}
E.~N{\"o}th, H.~Niemann, T.~Haderlein, M.~Decher, U.~Eysholdt, F.~Rosanowski, and T.~Wittenberg, ``Automatic stuttering recognition using hidden markov models,'' in \emph{Interspeech}, 2000.

\bibitem{barrett2022systematic-stutter1.0}
L.~Barrett, J.~Hu, and P.~Howell, ``Systematic review of machine learning approaches for detecting developmental stuttering,'' \emph{IEEE/ACM Transactions on Audio, Speech, and Language Processing}, vol.~30, pp. 1160--1172, 2022.

\bibitem{jouaiti2022dysfluency-stutter1.1}
M.~Jouaiti and K.~Dautenhahn, ``Dysfluency classification in stuttered speech using deep learning for real-time applications,'' in \emph{ICASSP}.\hskip 1em plus 0.5em minus 0.4em\relax IEEE, 2022, pp. 6482--6486.

\bibitem{bayerl2022detecting-stutter1.2}
S.~P. Bayerl, D.~Wagner, E.~N{\"o}th, and K.~Riedhammer, ``Detecting dysfluencies in stuttering therapy using wav2vec 2.0,'' \emph{arXiv preprint arXiv:2204.03417}, 2022.

\bibitem{zayats2016disfluency-stutter1.3}
V.~Zayats, M.~Ostendorf, and H.~Hajishirzi, ``Disfluency detection using a bidirectional lstm,'' \emph{arXiv preprint arXiv:1604.03209}, 2016.

\bibitem{montacie2022audio-stutter1.4}
C.~Montaci{\'e}, M.-J. Caraty, and N.~Lackovic, ``Audio features from the wav2vec 2.0 embeddings for the acm multimedia 2022 stuttering challenge,'' in \emph{Proceedings of the 30th ACM International Conference on Multimedia}, 2022, pp. 7195--7199.

\bibitem{kourkounakis2021fluentnet}
T.~Kourkounakis, A.~Hajavi, and A.~Etemad, ``Fluentnet: End-to-end detection of stuttered speech disfluencies with deep learning,'' \emph{IEEE/ACM Transactions on Audio, Speech, and Language Processing}, vol.~29, pp. 2986--2999, 2021.

\bibitem{alharbi2017segment-detection2}
S.~Alharbi, A.~J. Simons, S.~Brumfitt, and P.~D. Green, ``Automatic recognition of children’s read speech for stuttering application,'' in \emph{6th. Workshop on Child Computer Interaction (WOCCI 2017)}, 2017, pp. 1--6.

\bibitem{alharbi2020segment-detection3}
S.~Alharbi, M.~Hasan, A.~J. Simons, S.~Brumfitt, and P.~Green, ``Sequence labeling to detect stuttering events in read speech,'' \emph{Computer Speech \& Language}, vol.~62, p. 101052, 2020.

\bibitem{segment-detection4}
M.~Jouaiti and K.~Dautenhahn, ``Dysfluency classification in stuttered speech using deep learning for real-time applications,'' in \emph{ICASSP}, 2022, pp. 6482--6486.

\bibitem{shonibare2022frame-detection2}
O.~Shonibare, X.~Tong, and V.~Ravichandran, ``Enhancing asr for stuttered speech with limited data using detect and pass,'' \emph{arXiv preprint arXiv:2202.05396}, 2022.

\bibitem{wagner2024largelanguagemodelsdysfluency}
D.~Wagner, S.~P. Bayerl, I.~Baumann, K.~Riedhammer, E.~Nöth, and T.~Bocklet, ``Large language models for dysfluency detection in stuttered speech,'' in \emph{Interspeech}, 2024.

\bibitem{shih2024self-ssl-stutter}
Y.-J. Shih, Z.~Gkalitsiou, A.~G. Dimakis, and D.~Harwath, ``Self-supervised speech models for word-level stuttered speech detection,'' in \emph{2024 IEEE Spoken Language Technology Workshop (SLT)}.\hskip 1em plus 0.5em minus 0.4em\relax IEEE, 2024, pp. 937--944.

\bibitem{UDM}
J.~Lian, C.~Feng, N.~Farooqi, S.~Li, A.~Kashyap, C.~J. Cho, P.~Wu, R.~Netzorg, T.~Li, and G.~K. Anumanchipalli, ``Unconstrained dysfluency modeling for dysfluent speech transcription and detection,'' in \emph{2023 IEEE Automatic Speech Recognition and Understanding Workshop (ASRU)}, 2023, pp. 1--8.

\bibitem{lian-anumanchipalli-2024-towards}
J.~Lian and G.~Anumanchipalli, ``Towards hierarchical spoken language disfluency modeling,'' in \emph{Proceedings of the 18th Conference of the European Chapter of the Association for Computational Linguistics}, 2024, pp. 539--551.

\bibitem{Zhou2024YOLOStutterER}
X.~Zhou, A.~Kashyap, S.~Li, A.~Sharma, B.~Morin, D.~Baquirin, J.~Vonk, Z.~Ezzes, Z.~Miller, M.~Tempini, J.~Lian, and G.~Anumanchipalli, ``Yolo-stutter: End-to-end region-wise speech dysfluency detection,'' in \emph{Interspeech 2024}, 2024, pp. 937--941.

\bibitem{zhou2024stutter}
X.~Zhou, C.~J. Cho, A.~Sharma, B.~Morin, D.~Baquirin, J.~Vonk, Z.~Ezzes, Z.~Miller, B.~L. Tee, M.~L. Gorno-Tempini \emph{et~al.}, ``Stutter-solver: End-to-end multi-lingual dysfluency detection,'' in \emph{2024 IEEE Spoken Language Technology Workshop (SLT)}.\hskip 1em plus 0.5em minus 0.4em\relax IEEE, 2024, pp. 1039--1046.

\bibitem{ssdm}
J.~Lian, X.~Zhou, Z.~Ezzes, J.~Vonk, B.~Morin, D.~P. Baquirin, Z.~Miller, M.~L. Gorno~Tempini, and G.~Anumanchipalli, ``Ssdm: Scalable speech dysfluency modeling,'' in \emph{Advances in Neural Information Processing Systems}, vol.~37, 2024.

\bibitem{lian2024ssdm2.0}
J.~Lian, X.~Zhou, Z.~Ezzes, J.~Vonk, B.~Morin, D.~Baquirin, Z.~Mille, M.~L.~G. Tempini, and G.~K. Anumanchipalli, ``Ssdm 2.0: Time-accurate speech rich transcription with non-fluencies,'' \emph{arXiv preprint arXiv:2412.00265}, 2024.

\bibitem{OpenAI_chatgpt}
\BIBentryALTinterwordspacing
OpenAI, ``Chatgpt,'' 2022. [Online]. Available: \url{https://openai.com/chatgpt/}
\BIBentrySTDinterwordspacing

\bibitem{li2020universal}
X.~Li, S.~Dalmia, J.~Li, M.~Lee, P.~Littell, J.~Yao, A.~Anastasopoulos, D.~R. Mortensen, G.~Neubig, A.~W. Black \emph{et~al.}, ``Universal phone recognition with a multilingual allophone system,'' in \emph{ICASSP}.\hskip 1em plus 0.5em minus 0.4em\relax IEEE, 2020, pp. 8249--8253.

\bibitem{xu2021simple-w2v2-phoneme}
Q.~Xu, A.~Baevski, and M.~Auli, ``Simple and effective zero-shot cross-lingual phoneme recognition,'' \emph{Interspeech}, 2022.

\bibitem{wavlm-ctc}
``Wavlm-ctc-hugginface,'' \url{https://huggingface.co/microsoft/wavlm-large}.

\bibitem{graves2006connectionist-ctc}
A.~Graves, S.~Fern{\'a}ndez, F.~Gomez, and J.~Schmidhuber, ``Connectionist temporal classification: labelling unsegmented sequence data with recurrent neural networks,'' in \emph{International Conference on Machine learning}, 2006, pp. 369--376.

\bibitem{Baevski2020wav2vec2A}
A.~Baevski, H.~Zhou, A.~rahman Mohamed, and M.~Auli, ``wav2vec 2.0: A framework for self-supervised learning of speech representations,'' \emph{Advances in Neural Information Processing Systems,}, 2020.

\bibitem{chen2022wavlm}
S.~Chen, C.~Wang, Z.~Chen, Y.~Wu, S.~Liu, Z.~Chen, J.~Li, N.~Kanda, T.~Yoshioka, X.~Xiao, J.~Wu, L.~Zhou, S.~Ren, Y.~Qian, Y.~Qian, J.~Wu, M.~Zeng, X.~Yu, and F.~Wei, ``Wavlm: Large-scale self-supervised pre-training for full stack speech processing,'' \emph{IEEE JSTSP}, 2022.

\bibitem{Kouzelis2023WeaklysupervisedFA}
T.~Kouzelis, G.~Paraskevopoulos, A.~Katsamanis, and V.~Katsouros, ``Weakly-supervised forced alignment of disfluent speech using phoneme-level modeling,'' \emph{Interspeech}, 2023.

\bibitem{gorno2011classification-ppa}
M.~L. Gorno-Tempini, A.~E. Hillis, S.~Weintraub, A.~Kertesz, M.~Mendez, S.~F. Cappa, J.~M. Ogar, J.~D. Rohrer, S.~Black, B.~F. Boeve \emph{et~al.}, ``Classification of primary progressive aphasia and its variants,'' \emph{Neurology}, vol.~76, no.~11, pp. 1006--1014, 2011.

\bibitem{Enis2024FromLT}
\BIBentryALTinterwordspacing
M.~Enis and M.~Hopkins, ``From llm to nmt: Advancing low-resource machine translation with claude,'' \emph{ArXiv}, vol. abs/2404.13813, 2024. [Online]. Available: \url{https://api.semanticscholar.org/CorpusID:269292906}
\BIBentrySTDinterwordspacing

\bibitem{Kim2021ConditionalVA}
J.~Kim, J.~Kong, and J.~Son, ``Conditional variational autoencoder with adversarial learning for end-to-end text-to-speech,'' \emph{International Conference on Machine learning}, 2021.

\bibitem{zhou2024timetokensbenchmarkingendtoend}
\BIBentryALTinterwordspacing
X.~Zhou, J.~Lian, C.~J. Cho, J.~Liu, Z.~Ye, J.~Zhang, B.~Morin, D.~Baquirin, J.~Vonk, Z.~Ezzes, Z.~Miller, M.~L.~G. Tempini, and G.~Anumanchipalli, ``Time and tokens: Benchmarking end-to-end speech dysfluency detection,'' 2024. [Online]. Available: \url{https://arxiv.org/abs/2409.13582}
\BIBentrySTDinterwordspacing

\bibitem{cho2024jstsp}
C.~J. Cho, P.~Wu, T.~S. Prabhune, D.~Agarwal, and G.~K. Anumanchipalli, ``Coding speech through vocal tract kinematics,'' in \emph{IEEE JSTSP}, 2025.

\bibitem{wu23k_interspeech}
P.~Wu, T.~Li, Y.~Lu, Y.~Zhang, J.~Lian, A.~W. Black, L.~Goldstein, S.~Watanabe, and G.~K. Anumanchipalli, ``{Deep Speech Synthesis from MRI-Based Articulatory Representations},'' in \emph{Proc. INTERSPEECH 2023}, 2023, pp. 5132--5136.

\bibitem{lian22bcsnmf}
J.~Lian, A.~W. Black, L.~Goldstein, and G.~K. Anumanchipalli, ``{Deep Neural Convolutive Matrix Factorization for Articulatory Representation Decomposition},'' in \emph{Proc. Interspeech 2022}, 2022, pp. 4686--4690.

\bibitem{lian2023factor}
J.~Lian, A.~W. Black, Y.~Lu, L.~Goldstein, S.~Watanabe, and G.~K. Anumanchipalli, ``Articulatory representation learning via joint factor analysis and neural matrix factorization,'' in \emph{ICASSP 2023-2023 IEEE International Conference on Acoustics, Speech and Signal Processing (ICASSP)}.\hskip 1em plus 0.5em minus 0.4em\relax IEEE, 2023, pp. 1--5.

\end{thebibliography}

\end{document}